\documentclass{emulateapj}
\usepackage{aas_macros}
\usepackage{epstopdf}
\usepackage{epsfig}
\usepackage{natbib}
\usepackage{float}
\usepackage{gensymb}
\usepackage{amsmath}
\usepackage{lipsum}% http://ctan.org/pkg/lipsum
\usepackage{times}
\usepackage{comment}
\usepackage{color}
\usepackage{multirow}

\definecolor{purple}{RGB}{160,32,240}

\begin{document}

\title{Emission from the Ionized Gaseous Halos of Low Redshift Galaxies and Their Neighbors}
%the Second Halo Term}

\author{Huanian Zhang, Dennis Zaritsky, and Peter Behroozi \altaffilmark{1}}
\altaffiltext{1}{Steward Observatory, University of Arizona, Tucson, AZ 85719, USA; fantasyzhn@email.arizona.edu}

\begin{abstract}
Using a sample of nearly half a million galaxies, intersected by over 8 million lines of sight from the Sloan Digital Sky Survey Data Release 12, we extend our previous study of the recombination radiation emitted by the gaseous halos of nearby galaxies. We identify an inflection in the radial profile of the H$\alpha$+N[{\small II}] radial emission profile at a projected radius of $\sim 50$ kpc and suggest that beyond this radius the emission from ionized gas in spatially correlated halos dominates the profile. 
We confirm that this is a viable hypothesis using  results from a highly simplified theoretical treatment in which the dark matter halo distribution from cosmological simulations is straightforwardly populated with gas. Whether we fit the fraction of halo gas in a cooler (T $= 12,000$ K), smooth ($c = 1$) component (0.26 for galaxies with M$_* = 10^{10.88}$ M$_\odot$ and 0.34 for those with M$_* = 10^{10.18}$ M$_\odot$) or take independent values of this fraction from published hydrodynamical simulations (0.19 and 0.38, respectively), this model successfully reproduces the radial location and amplitude of the observed inflection. 
We also observe that the physical nature of the gaseous halo connects to primary galaxy morphology beyond any  relationship to the galaxy's stellar mass and star formation rate.
We explore whether the model reproduces behavior related to the central galaxy's stellar mass, star formation rate, and morphology. We find that it is unsuccessful in reproducing the observations at this level of detail and discuss various shortcomings of our simple model that may be responsible.

\end{abstract}


\section{Introduction}

Tracing the circumgalactic medium, CGM, is a long standing goal with a correspondingly extensive literature \citep[see][for a recent comprehensive review]{CGM2017}. The importance of the CGM arises from the expectation that it is the
reservoir for subsequent star formation in galaxies \citep{spitzer}, the depository for outflowing material, and the 
site of the majority of the baryons in galaxies \citep{Bregman,Werk2014}. 

We have recently turned a corner in our observational study of the CGM by advancing from studies that produced
ever decreasing upper limits on the resulting recombination emission line flux from 
this gas \citep{bland-hawthorn,dicaire,christlein,hlav,adams}
to detections of this emission utilizing either new techniques \citep[][hereafter Paper I]{zhang2016} or new instrumentation \citep{fumagalli,Bellhouse}. 

Of specific relevance to this study is our previous
detection of H$\alpha$+N[{\small II}] emission surrounding low redshift galaxies out to projected radii of 100 kpc, described in Paper I, using a sample of over 7 million lines of sight from SDSS DR12 \citep{SDSS12}. Using the same data source and technique,
we now turn to questions beyond  detection and basic characterization. Specifically, we explore the shape of the stacked emission line radial profile and  relationships between galaxy properties such as stellar mass, star formation rate (SFR), morphology, and environment and this radial profile. 
Our aim 
is to uncover variations among halo properties that can be used to identify the dominant ionization  source and to constrain the cold gas distribution, both of which will 
lead to a greater understanding of the physical conditions of the gas.

We identify an apparent inflection in the radial profile of the recombination line emission from halo gas (\S2). We confirm that this inflection is not due to uncertainties in the background measurement (\S2) and, using highly simplified models of the halo gas embedded within a distribution of dark matter halos produced from cosmological simulations, we demonstrate that the 
radial location and amplitude of the inflection is 
consistent with expectations arising from the consideration of the flux contributed by spatially correlated halos (\S3). 

We explore the models further by examining correlations between the radial emission line profiles and various central galaxy properties (\S2)
and expectations drawn from the models (\S3). Here,
we are less successful at reproducing the observations, but this disagreement likely reflect deficits in our simple models.
The comparison to our models is nevertheless helpful to guide intuition, but more physically realistic simulations are absolutely critical to achieving a robust and detailed understanding of the findings described here.
We adopt standard cosmological parameters $\Omega_m$ = 0.3, $\Omega_\Lambda =$ 0.7, $\Omega_k$ = 0 and the dimensionless Hubble constant $h = $ 0.7.

\section{Data Analysis and Results}
We follow the 
approach developed in Paper I.
We obtain spectra from the Sloan Digital Sky Survey Data Releases \cite[SDSS DR12]{SDSS12}, but now we  decrease the lower bound on the primary galaxy redshift range from 0.05 to 0.025 to increase the available number of lines of sight by roughly 1 million. Briefly, as described in Paper I, we classify galaxies that meet the criteria in redshift (now from 0.025 $< z < $ 0.2), luminosity ($10^{10}< L/L_\odot < 10^{11}$), and size (2 $< {\rm R_{50}}/{\rm kpc} <$ 10) as primary galaxies, and then study lines of sight to other galaxies that are within 1.5 Mpc projected separation of a primary galaxy and that were SDSS spectroscopic targets. For each such spectrum, we fit and subtract a 10th order polynomial to a 300 \AA\ wide section surrounding the wavelength of H$\alpha$ at the primary galaxy redshift to remove the continuum. 
We then measure the residual H$\alpha$ + N[{\small II}] flux within a velocity window corresponding to $\pm$275 km s$^{-1}$ from the primary galaxy
to capture the majority of the emission flux from the halo gas.  
To avoid having contamination by satellites of the target galaxy, we require the redshift difference between the primary and the line of sight target to be greater than 0.05 and we require that $\lvert f\rvert\le 0.3 \times 10^{-17}$ erg cm$^{-2}$ s$^{-1}$ \AA$^{-1}$, as in our previous work.
The closeness in wavelength of H$\alpha$ and N[{\small II}], our inability to identify the emission lines in individual spectra, and the unknown peculiar velocity of the halo gas relative to the primary galaxy,  preclude us from separating the flux from these lines in our measurement. We apply one final correction for a possible remaining systematic residual in the subtracted sky spectra by combining all of the spectra in the observed frame, fitting a 10th order polynomial, and subtracting the integral of that fit over the measurement window. 

\begin{figure}[!htbp]
\begin{center}
\includegraphics[width = 0.48 \textwidth]{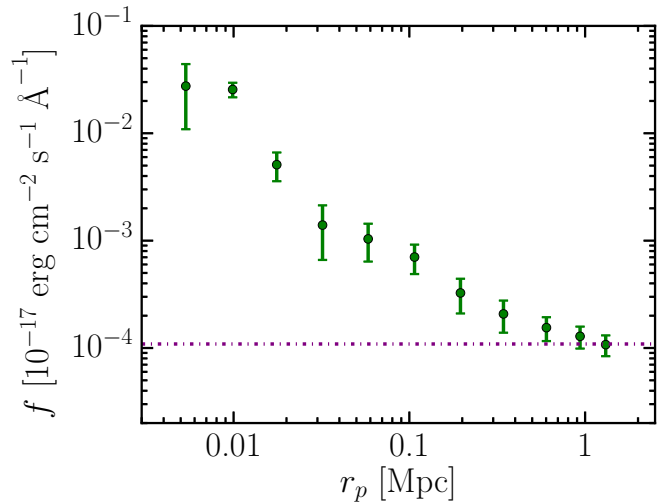}
\end{center}
\caption{The radial profile of H$\alpha$ + N[\small{II}] flux, $f$, for the entire data sample and the indication of the systematic limit on our detection based on the value obtained at large separation, $f_B$ (horizontal dotted line). Flux is significantly detected to beyond 100 kpc, but the character of the profile appears to change from steeply declining to moderately declining at $\sim 50$ kpc. We explore the significance of this claim and present our interpretation of this change of behavior in \S3.}
\label{fig:profile}
\end{figure}

Combining all of our final measurements from over 8 million lines of sight, sorted in radial bins, we obtain the radial profile of the H$\alpha$ + N[{\small II}] emission flux, $f$, around the average primary galaxy.\footnote{
The conversion factor to units between the values we present and those used commonly in the literature, erg cm$^{-2}$ s$^{-1}$ arcsec$^{-2}$, is 1.7.} The data
plotted in Figure \ref{fig:profile} represent the mean
values within each radial bin and the quoted uncertainty is the statistical uncertainty in that mean value derived from the dispersion of measurements in each bin. We present the plotted values in Table \ref{tab:data}.

\begin{deluxetable}{lr}
\tablewidth{0pt}
\tablehead{
\colhead{$r_p$} & \colhead{$f$} \\
\colhead {[kpc]} & {[$10^{-17}$ erg cm$^{-2}$ s$^{-1}$ \AA$^{-1}$]}
}
\startdata
5&  $(2.75\pm 1.65)\times 10^{-2}$ \\
10 & $(2.56\pm 0.40)\times 10^{-2}$ \\
18 & $(5.10\pm 1.51)\times 10^{-3}$ \\
32 & $(1.39\pm 0.71)\times 10^{-3}$ \\
58 & $(1.04\pm 0.39)\times 10^{-3}$ \\
107 & $(7.03\pm 2.15)\times 10^{-4}$ \\
195 & $(3.26\pm 1.16)\times 10^{-4}$ \\
343 & $(2.08\pm 0.68)\times 10^{-4}$ \\
602 & $(1.55\pm 0.38)\times 10^{-4}$ \\
936 & $(1.29\pm 0.30)\times 10^{-4}$ \\
1310 & $(1.07\pm 0.24)\times 10^{-4}$
\enddata
\label{tab:data}
\end{deluxetable}

A critical aspect of this analysis, in which we are attempting to extend our study to the largest possible projected radius, is the determination of the far-field value of $f$, or the ``background" level, $f_B$. In Paper I, we measured $f_B$ using the average of $f$ at large radii ($1 < r_p/{\rm Mpc} < 1.5$) and then presented and discussed the net emission flux, $\Delta f \equiv f - f_B$, as a function of projected radius, $r_p$. Here, because we are interested in the
shape of the outermost emission line flux radial profile and the possible flux contribution by the nearby environment, we are concerned that subtracting a single, uncertain $f_B$ value for all systems will inadvertently mask behavior of interest underneath a large uncertainty in the net emission flux.

To estimate the value of $f_B$ and its associated uncertainty, we first follow our approach from Paper I and measure $f_B$ from lines of sight at $ 1.0 < r_p/{\rm Mpc} < 1.5$, where we expect the emission to have dropped below our detection threshold. We measure $f_B = 1.09  \times 10^{-21}$ erg cm$^{-2}$ s$^{-1}$ \AA$^{-1}$. This non-zero result implies that we are either incorrect about our assumption that there is no detectable H$\alpha$+N[{\small II}] flux at such large separation or that there is a systematic uncertainty in our sky and continuum subtraction of comparable magnitude. To distinguish between these possibilities, we extend our examination by shifting the measurement window $\sim$ 100 \AA\ either blueward or redward, where there are no emission lines, and measure $f_B$ again from the lines of sight at large separation. For the blueward window shift we measure $f_B = 1.01  \times 10^{-21}$ erg cm$^{-2}$ s$^{-1}$ \AA$^{-1}$ and for the redward shift we measure $f_B = -0.85  \times 10^{-21}$ erg cm$^{-2}$ s$^{-1}$ \AA$^{-1}$. The discrepant results indicate that there is systematic error in $f_B$ at the level of $\sim \pm 1 \times 10^{-21}$ erg cm$^{-2}$ s$^{-1}$ \AA$^{-1}$ and so our initial measurement of $f_B$ provides no evidence of emission flux at $r_p > 1$ Mpc. We conclude that the measured $f_B$ represents a systematics-limited detection floor for our stacked lines of sight and so we do not subtract it.

In Figure \ref{fig:profile} we present the radial emission line flux profile from our stacked and radially binned measurements of $f$. In addition, we highlight the detection floor attributable to systematic uncertainties in $f_B$. We summarize the findings so far as 1) there is detectable emission, well above the $f_B$ limit, out beyond 100 kpc, 2) there is a steep decline in the profile from the central galaxy to $\sim 50$ kpc, and 3) there is an intermediate region from $\sim$ 50 kpc extending beyond 100 kpc where the profile flattens but remains above $f_B$. 
The inflection between the steep inner decline and the flatter outer region is what we consider to be evidence for emission from spatially correlated 
halos and what we discuss in more detail in \S\ref{sec:theory}. If the reader is somewhat skeptical of our claim that the profile changes character at $r_p \sim$ 50 kpc, we request their indulgence until we present the calculated emission profile from cold gas in an isolated halo. 

\subsection{The Emission Line Profile and Primary Galaxy Properties}

To explore the dependencies of the H$\alpha$ + N[{\small II}] emission flux on galaxy properties further  than we did in Paper I, we extract measures of the galaxy's Sersic index ($n$) 
and absolute magnitude ($M$) from \citet{simard}, stellar mass (M$_*$) from \citet{Kauffmann1,Kauffmann2} and \citet{Gallazzi}, and current star formation rate (SFR) from \citet{Brinchmann}.

\subsubsection{Dependence on Stellar Mass}

We begin our exploration by considering variations with stellar mass. A natural expectation is that  more massive galaxies have more halo gas and therefore a brighter and more extended emitting halo. 
To test this expectation, we divide the primary galaxies into two nearly equal sized subsamples by setting the demarcation at a stellar mass of $10^{10.45} {\rm M}_{\odot}$. 
To quantify any differences between the two samples we compare the mean values of $f$ in unit of 10$^{-17}$ erg cm$^{-2}$ s$^{-1}$ \AA$^{-1}$ at radii of 10, 50, and 300 kpc (Table \ref{tab:data}).

\begin{deluxetable}{lrr}
\tablewidth{0pt}
\tablehead{
\colhead{$r_p$} & \colhead{$f$ (M$_* > 10^{10.45}$M$_\odot$)} & \colhead{$f$ (M$_* \le 10^{10.45}$M$_\odot$)} \\ 
\colhead {[kpc]} & \multicolumn{2}{c}{[$10^{-17}$ erg cm$^{-2}$ s$^{-1}$ \AA$^{-1}$]}}
\startdata
10&  $(1.38\pm 0.49) \times 10^{-2}$ & $(3.79\pm 0.61) \times 10^{-2}$ \\
50 &  $(1.04\pm 0.60) \times 10^{-3}$  & $(0.54\pm 0.48) \times 10^{-3}$ \\
300 & $(3.13\pm 0.98) \times 10^{-4}$ & $(2.33\pm 1.12) \times 10^{-4}$
\enddata
\label{tab:data}
\end{deluxetable}

The mean $f$ values show some statistically significant differences between mass samples, but only at the smallest radii and not perhaps in the expected sense. It is the low stellar mass galaxies that show the larger emission fluxes in the 10 kpc bin. We interpret this result to mean that late-type, star forming galaxies, which are generally of lower stellar mass, have more luminous
halos at these radii. By 50 kpc the fluxes are nearly indistinguishable, suggesting whatever connection existed between the emission and primary galaxy is absent at these, and larger, radii. 
We draw two conclusions from this initial comparison. First, significant and interesting differences in halo emission line properties are detectable among galaxy subsamples. Second, although we selected by stellar mass, the differences are likely to be tied to a variety of factors that have different levels of influence as a function of $r_p$ and may be difficult to unravel if one cannot control for these factors. 

\subsubsection{Dependence on SFR}

One of our goals is to identify the source(s) of ionizing photons for this emitting gas. 
Escaping high energy photons from the primary galaxy are one possibility.
If so, the
strength of the measured emission flux could correlate with the star formation rate. Furthermore, such 
a scenario could explain our previous result in which the 
more massive galaxies, which are likely to be dominated by earlier types, have less emission flux at small $r_p$ than do the lower mass galaxies, which are likely to be dominated by later types. Quantifying such a correlation could lead to a measurement of the escape fraction, which would in and of itself be a significant and valuable result.

To search for  
a connection between the emission line profile and the SFR of the primary galaxy, we divide the sample nearly equally
into a low SFR subsample (SFR $<$ 0.70 M$_\odot$/year) and a high SFR one (SFR $>$ 0.70 M$_\odot$/year). The demarcation is chosen only to balance the size of the subsamples, not for any physically motivated reason. 

\begin{figure*}[!htbp]
\begin{center}
\includegraphics[width = 0.8 \textwidth]{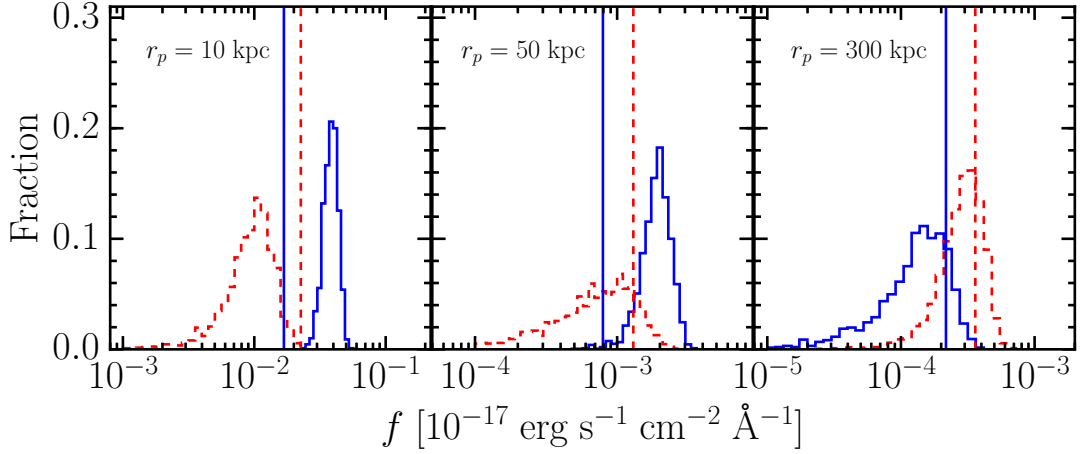}
\end{center}
\caption{H$\alpha$+ N[II] emission line signal strength at different radii for stellar mass matched realizations of the high SFR (blue solid histogram) and low SFR galaxy samples (red dashed histogram). Vertical lines represent results of models described in \S\ref{sec:theory} for the mean of stellar mass matched high SFR (blue solid line) and low SFR samples (red dashed line).
}
\label{fig:massmatchedSFR}
\end{figure*}

We now know that stellar
mass might also be a physical driver of differences in the flux profiles, so we control for  stellar mass differences.
We construct stellar mass-matched subsamples sorted by SFR by first
binning the low and high SFR subsamples by stellar mass. For each stellar mass bin, we include all of the primary galaxies in whichever SFR subsample has fewer galaxies and 
randomly select an equally sized sample from the other set. 
We do this for all of the mass bins to obtain one realization of mass matched low and high SFR subsamples
and measure the average of $f$ in bins centered at $r_p = $ 10, 50, and 300 kpc.
The whole process is repeated 2000 times to obtain the distributions of $f$ at the three different radii presented in 
Figure \ref{fig:massmatchedSFR}. 

The results highlight three qualitatively different radial regimes. At small $r_p$, 10 kpc,
the emission in the high SFR subsample is a few times higher than that in the low SFR subsample.
We interpret this as the result of the higher star formation rate and suggest that it arises from a combination of a stronger
ionizing radiation field generated by the star formation and a more massive cold gas reservoir. Although these are plausible explanations, physical reality is likely to be 
complicated by the detailed internal geometry of the central galaxy. Studies of the
Ly$\alpha$ emission from starbursting systems have highlighted the complex role of dust content
and geometry on the escaping radiation
\citep{LARS2013,LARS2014,LARS2017}. 
At intermediate $r_p$, 50 kpc, the emission flux is becoming more similar in the two subsamples and therefore less dependent on the star formation rate of the central galaxy. We interpret this result as indicating that any escaping radiation has a declining effect, for normal low z galaxies, on the ionization state of the halo gas beyond several tens of kpc. A corollary to this inference is that the ionization at this and larger radius is, to a greater degree, due to other sources of ionization such as the intergalactic radiation field or to shock heating. Finally, and perhaps somewhat puzzling at first is the reversal seen at the largest $r_p$, 300 kpc, where the emission flux is greater in the subsample of primaries with lower SFR. 

We conclude that in mass-controlled subsets, we find that the emission line fluxes of galaxies at $r_p \sim 10$ kpc are strongly related to the star formation rate of the primary galaxy, but that beyond $\sim$ 50 kpc any effects of the primary galaxy are weaker and that differences, where they exist, are likely due primarily to external agents. 

\begin{figure*}[!htbp]
\begin{center}
\includegraphics[width = 0.8 \textwidth]{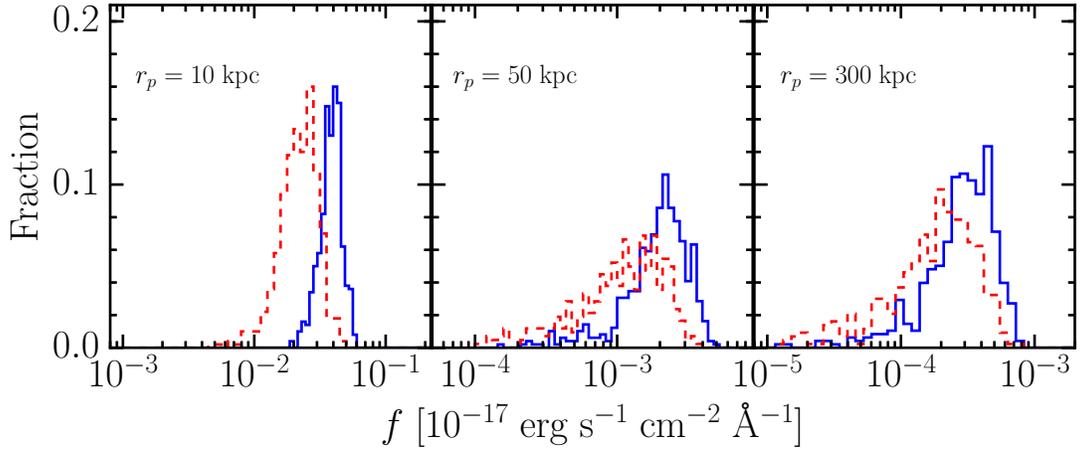}
\end{center}
\caption{H$\alpha$+ N[II] emission line signal strength at different radii for stellar mass and morphology matched realizations of the high SFR (blue solid histogram) and low SFR galaxy samples (red dashed histogram).
}
\label{fig:bothmatchedSFR}
\end{figure*}

Next, we control for differences in morphology, in addition to controlling for stellar mass.
As a proxy for morphology we use the Sersic index \cite[]{sersic}, classifying those primaries with $n > 3$ as spheroids and $n \le 2$ as disk.  
We present the $f$ distributions at the three selected radii in 
Figure \ref{fig:bothmatchedSFR}.

The pattern of results is somewhat similar to the earlier one, with the more strongly star forming galaxies having larger fluxes at 10 and 50 kpc, although at 300 kpc the ordering changes relative to the previous results. 
We conclude that 1) the local properties that drive the emission (local ionization field, halo gas density distribution, gas temperature) are different both in the halo of the primary galaxy and in the local environment for high and low star forming galaxies and 2) primary galaxy morphology contains additional information about the physical nature of the CGM beyond that arising from its relationship to the star formation rate. 

\section{Theoretical Models and Discussion}
\label{sec:theory}

We base our modeling on halo merger trees from the \textit{Bolshoi-Planck} simulation \citep{Klypin16,RP16}, a dark matter-only simulation with comoving side length of 250 Mpc $h^{-1}$, mass resolution of $1.8 \times 10^{8} M_\odot$ (2048$^3$ particles), and force resolution of $1$ kpc $h^{-1}$.  Halos were found with the \textsc{Rockstar} phase-space halo finder \citep{BehrooziRockstar}, and merger trees were generated with the \textsc{Consistent Trees} code \citep{BehrooziTrees}.  The simulation adopted a flat, $\Lambda$CDM cosmology with $\Omega_M = 0.307$, $\Omega_B = 0.048$, $h=0.678$, $n_s = 0.96$, and $\sigma_8 = 0.823$, similar to recent \textit{Planck} constraints \citep{Planck2013}.

Galaxy stellar masses and star formation rates were modeled with the \textsc{UniverseMachine} code (P.\ Behroozi et al., in prep.).  Briefly, the code empirically determines how galaxy star formation rates should depend on host dark matter halo mass, halo accretion rate, and redshift so that many observables are reproduced.  For this project, we used catalogs at $z=0.1$ (the median of the observed target galaxies) that match observed galaxy stellar mass functions, specific star formation rates, quiescent fractions, and correlation functions (split by color, mass, and redshift).  Because the code does not currently output galaxy $r$-band luminosities or sizes, we matched mock galaxies to the closest SDSS galaxies in stellar mass and star formation rate.  The mock catalogs thus reproduce by definition the joint probability distribution for $r$-band luminosity and galaxy size given galaxy stellar mass and star formation rate.

Once we have a catalog of halos and the galaxies associated with them, we  populate each halo with gas.
Within each NFW dark matter halo \citep{navarro} we  adopt a hydro-equilibrium isothermal model as the description of the gas mass density profile:
\begin{equation}
	\rho_g(r) = h_f ~ \rho_0~ {\rm exp}{\Big[-\Delta_{NFW}\Big(1 - \frac{{\rm ln}(1+r/r_s)}{r/r_s}\Big)\Big]}
\end{equation}
where $\Delta_{NFW} = 4\pi {\rm G} \rho_0r_s^2\mu m_p/({\rm k}_{\rm B} {\rm T}_{\rm vir})$, $h_f$ is the hydrogen gas mass fraction, $\mu$ is mean molecular weight, $m_p$ is the proton mass and T$_{\rm vir}$ is the virial temperature \citep{Capelo2010}. To estimate $h_f$, we adopt the cosmological baryon fraction, $\simeq 0.16$ \citep{Planck2013}, multiply it by the fraction of baryons in a galaxy that are in the halo as opposed to being in the stars and interstellar medium of the central galaxy
\citep[$\sim 0.85$;][]{Behroozi2010,McGaugh2010,CGM2017} and then by the fraction of that baryonic mass that is in hydrogen atoms, $\sim 0.75$.
The dominant uncertainty in this calculation comes  from the adopted fraction of the baryons that are in the halo, which could be as low as 0.6 but probably not 
much higher than 0.85 \citep[see][and references therein for a recent baryon accounting in the Milky Way]{zaritsky17}.
For primordial gas,
$\mu$ is $\sim$ 0.59 because metals are trace elements and the hydrogen/helium mass ratio is 3:1. Even though the halo is metal enriched \citep{CGM2017}, the mass fractions are nearly unchanged and uncertainties in these values are not the dominant ones in our calculation. We calculate the virial temperature from the halo virial velocity, $v_{\rm vir}$, using

\begin{equation}
	{\rm T}_{\rm vir} = \frac{\mu m_p}{2{\rm k}_{\rm B}} v^2_{\rm vir} \simeq 3.6 \times 10^5 {\rm K} ~\Big(\frac{v_{\rm vir}}{100 ~ {\rm km/s}}\Big)^2 
\end{equation}
\noindent
Finally, we calculate the H$\alpha$ recombination rate in unit of cm$^3$ s$^{-1}$ for gas with temperature T in unit of $10^4$K and density $\rho_g$ using the fitting function for the recombination rate,
\begin{equation}
\alpha = 10^{-13} \frac{2.274 ~{\rm T}^{-0.659}}{1+1.939 ~{\rm T}^{0.574}},
\end{equation}
\noindent
from \cite{Pequignot}.

To create mock observations, we select a sample of primary galaxies.
We first randomly select lines of sight through the simulation volume that are presumed to be aimed at a background source that SDSS would have targeted. Cataloged halos within a projected separation of 1 Mpc from this line of sight are considered as hosts of potential primary galaxies, even if they are satellites within a larger halo. If the hosted galaxies meet the luminosity and size requirements ($10^{10} < L/L_\odot < 10^{11}$ and 2 $< {\rm R_{50}}/{\rm kpc} <$ 10) and if their apparent magnitude are above the SDSS targeting limit then they are accepted as primary galaxies. 
In detail, we use the observed galaxy luminosity-redshift relation of our primary galaxies to define the limiting luminosity at each redshift and then draw from the model to match that distribution. 
Because the mock catalog uses the luminosity-size distribution of the SDSS to set sizes, the joint luminosity-size distribution in the mock catalogs matches the SDSS by construction. 
Subsequently,  
any cataloged halo whose projected separation from the primary galaxies, $r_p$, is less than its virial radius, $r_{\rm vir}$, and whose recessional velocity is within 275 km s$^{-1}$ of the primary galaxy's recessional velocity is an associated halo that could possibly be contributing H$\alpha$ flux to the line of sight spectrum. For computational efficiency we neglect 
contributions from halos with $r_p > r_{\rm vir}$. 
We run some realizations where we change this criteria to $r_p > 5 r_{\rm vir}$ and find no detectable difference in the results. The SDSS magnitude limit directly affects the primary halo selection function, but not the  associated halo selection function, other than through correlations that may exist between the properties of  associated halos and the primary ones. 

\begin{figure}[!htbp]
\begin{center}
\includegraphics[width = 0.48 \textwidth]{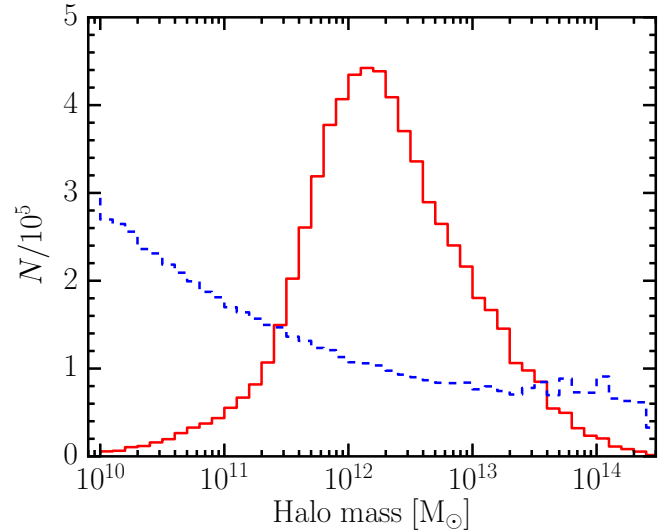}
\end{center}
\caption{The halo mass distribution of the primary (solid red line) and secondary (dashed blue line) halos. The halo mass of the target galaxy is mostly limited to $10^{11} < {\rm M/M}_\odot < 10^{13}$ because of the luminosity and half light radius selection criteria. The secondary halos are more representative of the full cosmological halo sample.}
\label{fig:halo_mass}
\end{figure}

We present the resulting model halo mass distributions for both the primaries and the associated halos in Figure \ref{fig:halo_mass}. The halo mass of the primary galaxy is constrained within the range $10^{11} - 10^{13}$ M$_\odot$, reflecting the luminosity and half light radius criteria we apply. The associated halo mass distribution is more representative of the universal halo distribution, with a preponderance of low mass halos. The mean stellar mass, M$_*$, for the primary halos is $10^{10.88}$ M$_\odot$ and
is $10^{10.18}$ M$_\odot$ for the associated halos.

Once we have our sample of primary and associated halos, we then integrate the square of the gas density, $\rho_g^2(r)$, projected along the line of sight, combine that measurement with the recombination rate, and derive a predicted H$\alpha$
flux measurement. We truncate the density profile of each associated halo at its virial radius for computational efficiency. We find no significant difference in our results in a test run where we
truncate the profiles at $5 r_{\rm vir}$. We calculate the flux
for a 2$^{\prime\prime}$ fiber (the BOSS fiber size) on the sky. 

One significant difference between the data and the simulated measurements is that the observations include the contribution of N[{\small II}]. To account for this difference, 
we measure the N{[\small II]} contribution from the stacked spectra for $r_p <$ 50 kpc, where the lines are of sufficiently high S/N that they are well resolved, and measure that the H$\alpha$ flux is 0.68 of the combined flux. Therefore, for our data and model comparisons, we multiply the modeled fluxes by 1.46 to account for N[{\small II}] emission. The H$\alpha$/N[{\small{II}}] ratio can change as a function of radius \citep{miller}, but this is 
beyond the scope of our models and the introduced
uncertainty is likely to be lower than that 
introduced by other simplifying assumption (see \S3.3).

\subsection{Single Temperature Component}

\begin{figure}[!htbp]
\begin{center}
\includegraphics[width = 0.48 \textwidth]{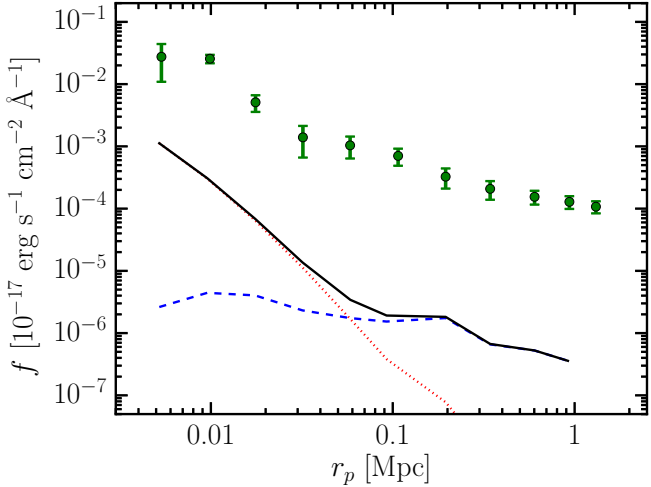}
\end{center}
\caption{Comparison of data to simulation with all halo gas at the corresponding halo virial temperature. The dotted red line represents the H$\alpha$+N[II] emission arising in the primary halo of the target galaxy and the dashed blue line from the associated halos. The solid black line represents the combined emission in the models. The green circles with error bars represent our measurements (Table \ref{tab:data}). The predicted emission from this model is orders of magnitude lower than the data and even below the background level, $f_B$, at most radii.}
\label{fig:Tvir}
\end{figure}

The recombination rate is temperature sensitive, so assumptions regarding the physical state of the gas drive the resulting emission measurements. We begin our first model by stipulating that the gas component is at a single temperature in each halo and that the temperature is the virial temperature. 
Because the typical virial velocity of a primary galaxy in our sample is $\sim$ 100 km/s or larger, the virial temperature is $\sim$ $10^{5}$ to $10^{6}$ K. Hydrogen recombination at such high temperatures is highly suppressed, resulting in simulated emission fluxes that are orders of magnitude lower than those observed (Figure \ref{fig:Tvir}). We conclude that either the observed emission has a completely different origin or our first model does not represent reality. This result is not surprising given the 
existing evidence for a multiphase CGM \citep{CGM2017} and it strongly motivates a modification to our simple model. Because the shape of the theoretical radial emission profile is qualitatively similar to that observed, we are encouraged and suggest that 
only a simple modification is needed, presumably related to the assumption of a single high temperature component.

\subsection{A Two Temperature Component Model}

Various absorption line studies \citep{Werk2014,CGM2017}, theoretical models \citep{keres,Ford2014}, and our own work in Paper I
conclude that much of the halo gas is at significantly lower temperatures than the virial temperature. 
In Paper I, we inferred that the emitting gas temperature is $\sim$ 12,000 K. Therefore,  we now
explore a model where the gas is in two phases: one at the virial temperature and another at 12,000 K.  While we have a mechanism for computing the radial density profile of gas at the virial temperature within a dark matter halo, we do not have a simple way to estimate the corresponding distribution of 10$^4$ K gas embedded within the hotter gas. We will assume that a fraction $C_f$ of the gas is in the cold component and that it follows the same radial profile as the hotter gas. The last potential complication in our simple model is whether this gas is clumped. If it is clumped, then the higher local density knots will emit more recombination radiation than if that same quantity of gas is distributed smoothly. For now we will assume a smooth distribution, but we will return to this issue later. 

\begin{figure}[!htbp]
\begin{center}
\includegraphics[width = 0.48 \textwidth]{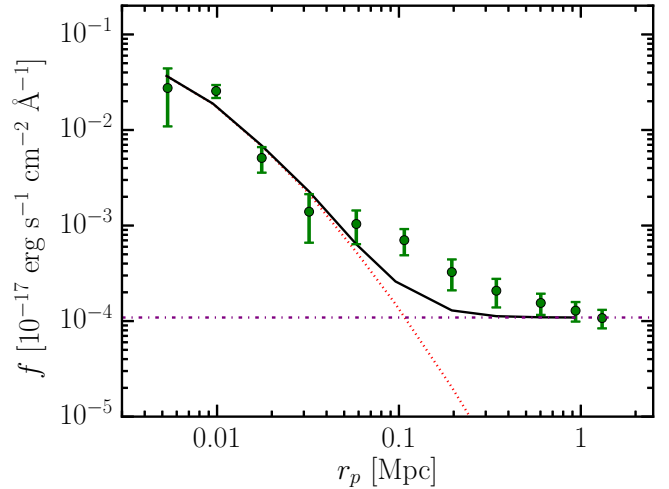}
\end{center}
\caption{
Comparison of data and simulation with fraction, $C_f$, of the halo gas at 12,000 K set to 0.27 and the remainder at the corresponding halo virial temperature for all halos. In this Figure we show only the H$\alpha$+N[II] emission arising from the primary halo (dotted red line) and the systematic limit on our measurement of the background, $f_B$ (horizontal line). Together (solid black line), these two components match the observed emission profile (green circles with error bars) at small and large $r_p$, but fail at intermediate $r_p$, illustrating the need for an additional component.
}
\label{fig:Tconst_fB}
\end{figure}

In Figure \ref{fig:Tconst_fB} we show our first attempt at fitting such models.
We consider only the contribution from the primary galaxy halo itself, which is indicated by the red dotted line for an adopted $C_f = 0.27$, and the contribution from the background, $f_B$. We determine  $C_f$ by minimizing $\chi^2$. The model fits the data well out to $\sim 50$ kpc and at $r_p > 1$ Mpc. Within 50 kpc, the one outlier is the second innermost point, which at $r_p \sim 10$ kpc is perhaps unlikely to be well described with this simple model. Only considering the data out to $\sim 50$ kpc, $\chi^2 = 7.02$, for which the model cannot be ruled out with greater than 80\% confidence. 
However, the $\chi^2$ value for $r_p < 1$ Mpc is 17.87, for which the model can be ruled out with 97\% confidence.

The model principally fails to reproduce the fluxes observed at intermediate radii. These fluxes are significantly larger than the background uncertainty and cannot be reproduced by modeled emission from the central halo. Increasing $C_f$ would raise the entire red curve, thereby grossly overproducing the flux at $r_p < 50$ kpc. Models with a single halo component plus a systematic background floor cannot reproduce the observations.

\begin{figure*}[!htbp]
\begin{center}
\includegraphics[width = 0.8 \textwidth]{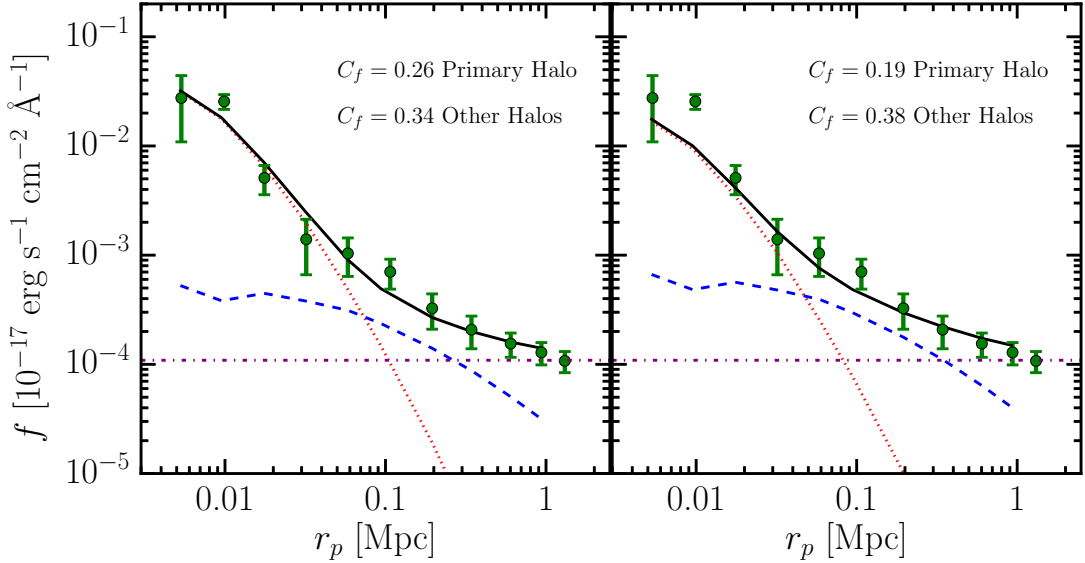}
\end{center}
\caption{ Comparison of data to simulations. Models assign different $C_f$'s to primary and  associated halos, but all have the smoothly distributed cold halo gas temperature set at 12,000 K and the remainder of the gas at the corresponding halo virial temperature for all halos. In this Figure we show the H$\alpha$+N[II] emission arising from the primary halo (dotted red line) and the  associated halos (blue dashed line) and the systematic limit on our measurement of the background, $f_B$ (horizontal line).
Together (solid black line), these three components are compared to the observed emission profile (green circles with error bars).
The left panel shows model results from our minimization of $\chi^2$ while varying $C_f$'s. Right panel shows the model results when we adopt the $C_f$ from the
relation between $C_f$ and stellar mass produced by independent hydrodynamical simulation \citep{Ford2014}.}
\label{fig:Tconst}
\end{figure*}

In Figure \ref{fig:Tconst} we present the next 
iteration of the models, now including the contribution from correlated, nearby halos. Because the primary and associated halos have a different mass distribution, we expect the characteristic $C_f$'s to differ. Simulations of the halo gas show such a variation with the galaxy's stellar mass \citep{CGM2017}. The average stellar mass for our modeled primary galaxies is $10^{10.88} {\rm M}_\odot$ and is $10^{10.18} {\rm M}_\odot$ for our secondary galaxies. Therefore, in our fitting we allow for different $C_f$ values for the primary and secondary halos.
Our best fit values are $C_f = 0.26$ and 0.34 for the primary and secondary halos, respectively, and the fit (left panel of Figure \ref{fig:Tconst}) is excellent. Alternatively, we can 
adopt $C_f$ from independent hydrodynamical simulations of galaxy halos \citep{Ford2014}, which provide a relationship between $C_f$ and stellar mass. That relation predicts that the appropriate $C_f$'s are 0.19 and 0.38, respectively. Those values also produce a good fit, and have the added virtue of removing all fitting freedom.

With or without fitting $C_f$'s, we now obtain a good match to the observations and conclude that the emission profile inflection at $r_p \sim 50$ kpc can be explained as the onset  of dominance by associated halos.
The best fit matches the observations quite well with a  $\chi^2$ of 9.17 for $r_p < 1$ Mpc.
The hypothesis that the decrease in $\chi^2$, from 17.87 to 9.17, resulting from the addition of the added freedom of allowing a second component, is consistent with random chance can be rejected with 99.8\% confidence. We conclude with high
confidence that there is evidence for the  contribution from associated halos. 

\begin{figure*}[!htbp]
\begin{center}
\includegraphics[width = 0.8 \textwidth]{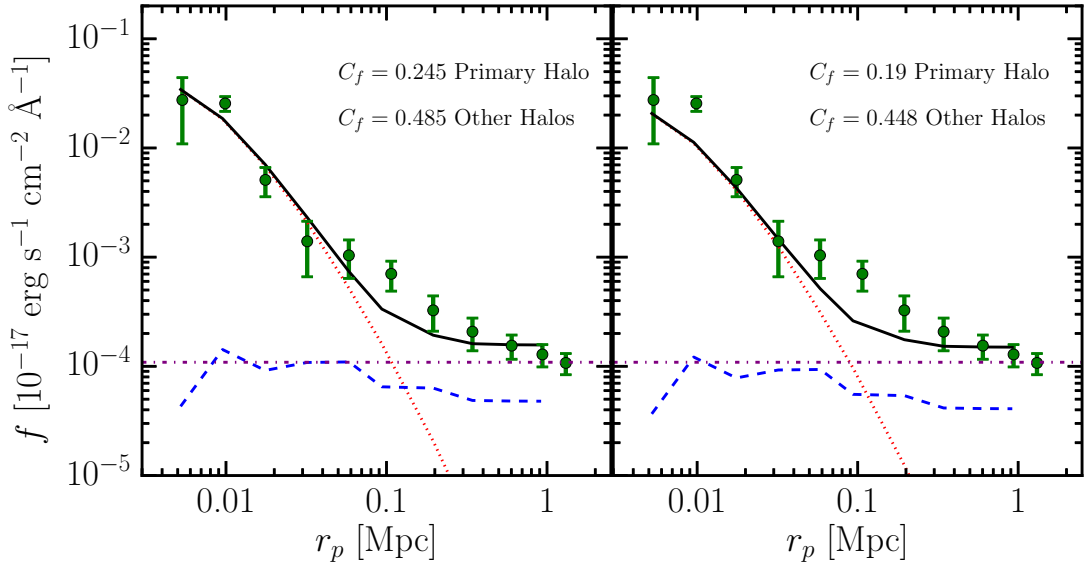}
\end{center}
\caption{ Comparison of data to simulations  that exclude primary galaxies that are satellites within larger halos. We refit $C_f$'s for primary and secondary halos. In this Figure we show the H$\alpha$+N[II] emission arising from the primary halo (dotted red line) and the secondary halos (blue dashed line) and the systematic limit on our measurement of the background, $f_B$ (horizontal line).
Together (solid black line), these three components are compared to the observed emission profile (green circles with error bars).
The left panel shows model results from our minimization of $\chi^2$ while varying $C_f$'s.
Right panel shows the model results when we adopt the $C_f$ from the
relation between $C_f$ and stellar mass produced by independent hydrodynamical simulation \citep{Ford2014}.}
\label{fig:Tconst_NOsate}
\end{figure*}

At this point, a natural question regards the nature of these associated halos. Are they principally satellites of the primary, parents of the primary, or distinct second halos? For various reasons we consider satellites as an unlikely significant source of H$\alpha$ flux. First, the SDSS target itself could not be a satellite because of our redshift difference requirement, which corresponds to a line-of-sight difference of $\sim$200 Mpc. Second, the odds that satellites are interlopers along the lines of sight are low because of their small cross sections. Third, we have eliminated lines of sight with significantly strong $f$ and therefore a few, strongly emitting satellites could not be the source of the signal. Finally, most satellites within 100 kpc of $L_*$ galaxies are gas poor. The models confirm this conclusion with virtually no flux originating in halos of $M < 2.5\times 10^{11} \rm M_\odot$, a limit that is ten times the resolution limit of the \textsc{UniverseMachine} mock catalogs.   Considering the second option, we find
that about 20\% of the primary galaxies in the simulations are satellite galaxies within group and cluster size halos. In our current model, these parent halos contribute significantly to the excess flux above the single halo expectation at intermediate radii (50-100 kpc), as we show in Figure \ref{fig:Tconst_NOsate}. Refitting $C_f$ for the case with no satellites results in values of $C_f > 0.4$ for the associated halos, but does not reproduce the flux at these intermediate radii.
Lastly, considering the third option, we find that distinct second halos contribute a modest amount, although that amount might be critical to explaining the flux detected at $r_p > 200$ kpc. Obviously, we are approaching the limits of the current measurements, but the models do suggest that the second halo term could be detected. Definitively measuring the second halo contribution will require improved precision.

So far, our simple model seems sufficient to explain the observations. However, there are clear complications that must be present. First, we have adopted a single temperature for the cold component, not just from halo to halo but also radially within each halo. 
The recombination rate for the adopted
temperature of 12,000 K is $6.40 \times 10^{-14}$ cm$^3$ s$^{-1}$
but changes nearly inversely with temperature, ranging from $1.30 \times 10^{-13}$ cm$^3$ s$^{-1}$ at 6000 K, to $3.04 \times 10^{-14}$ cm$^3$ s$^{-1}$ at 24000 K. This dependence means that relatively modest, factor of 2, discrepancies in emission flux
could be explained by plausible temperature variations within what we refer to as the cold
component.

Second, we have adopted a model with a smooth density distribution of cold gas. The cold gas is likely to be clumped and the emission flux depends on density squared. Stacking a large number of systems is presumably helping to average over the most discrepant cases, but a systematic bias in the modeled fluxes, depending on the fraction of  gas that is in clumps, could be present.

To allow for more generality in our discussion, we define an emission factor, $\epsilon = C_f({\rm T}/12000 {\rm K})^{-0.5}(c/1)$, where $c$ is a gas clumping factor.  Our constraints above are in practice on the emission factor $\epsilon$, and the cold fraction $C_f$ could vary as long as it is balanced by corresponding changes in the gas temperature $T$ and/or the clumping factor $c$ from our assumptions above. 

Third, our model makes no attempt to distinguish between primary galaxies that are the dominant galaxy in their halo versus those that are satellites within a more massive halo. It is entirely reasonable to expect the satellite galaxies to have different gaseous halo properties than the isolated ones. To first order, if we presume that the satellites have lost their gaseous reservoir that would simply depress the predicted $f$ by the fraction of satellites in the sample, $\sim 20\%$. Such a difference could easily be compensated in the models by a commensurate increase in the clumping discussed previously. As such, even though this discussion clearly points to a shortcoming of our models, the effect is likely to be well within our rather large uncertainties.

The effect of nearby halos on measurements of halo gas has been explored before in theoretical work. 
For example, \cite{Nelson2017} explored the contributions from other halos to the column density of absorption line systems arising from highly ionized oxygen O{\small VI}, O{\small VII}, and O{\small VIII} gas. They found in their simulations that the contribution from second halos dominate in the 3D profiles 50 to 100 kpc beyond $r_{\rm vir}$ for halos of $10^{12} M_\odot$. 
A comparison between emission and absorption line properties of different tracers holds promise as a powerful model discriminant because of the different dependencies of the two measurements on the temperature and density of the gas.

The need for a more complex model can be seen clearly when we examine our model further. In the case with no fitting, we adopted the $C_f$ from \cite{Ford2014} corresponding to the mean stellar mass of the associated halos (log $({\rm M}_*/{\rm M}_\odot) =10.18$). However, in that model the {\sl halo-emission, flux-weighted} mean stellar mass, which might be the appropriate one to use because the bulk of the flux is coming from such systems, is significantly higher (log $({\rm M}_*/{\rm M}_\odot) = 10.72$). For this high a stellar mass the \cite{Ford2014} study predict a much lower $C_f$. This choice of $C_f$ leads to a much lower flux from the associated halos, which results in an emission profile that does not match that observed. Given this 
conundrum, we also ran a model where the $C_f$ of each individual associated halo was adjusted using the \cite{Ford2014} relation. This model also failed to reproduce the radial profile at intermediate radii. 

There are two ways to resolve 
this discrepancy. First, the \cite{Ford2014} relation between M$_*$ and $C_f$ may be incorrect. This is an interesting possibility in that our observations may be telling us how the hydrodynamical models need to be improved. Second, and perhaps even more likely, discrepancies may be due to variations in $c$ and T as a function of M$_*$ (as well as radius and environment) that are not accounted for in our simple model. Again, this may lead to greater insight into the CGM, particularly if some of the degeneracies involved in calculating the emission fluxes can be broken by including absorption line constraints.   

\subsection{Higher Order Tests of the Model}
\label{gal_props}

So far, we have demonstrated that the simple fitted model matches the data from the full sample and reveals that the galaxy's neighboring halos are responsible for the dominant contribution to H$\alpha$ + N{[\small II]} emission at intermediate $r_p$. We now investigate what the model predicts about how the primary galaxy's properties correlate to the emission profile of the gaseous halo and how those predictions compare to the observations.

For the comparison of the halo properties as a function of the SFR of the primary galaxy, we divide the entire model sample into high SFR subsample (SFR $> 0.7 {\rm M}_\odot$/year) and low SFR subsample (SFR $\le 0.7 {\rm M}_\odot$/year) and follow the same
mass matching procedure as we did
for the observational sample. 
Rather than showing the resulting $f$ distribution, as we do for the data, for the models we only show the resulting mean values in Figure \ref{fig:massmatchedSFR}. 

There are some successes in the model predictions, for example the agreement between the predicted and observed $f$ values for the high and low SFR galaxies at large $r_p$. However, the failures may be more illuminating. The models fail for both the high and low SFR galaxies at small $r_p$, by underpredicting and overpredicting the emission flux respectively. In essence, the model predicts that the halo emission from high and low SFR galaxies is much more similar at these radii than we observe it to be. 
Clearly, there are 
important physical details, the role of feedback and the nature of escaping ionizing radiation, that are not included in our simple models. One can imagine that feedback has removed at least some of the gas reservoir in early type galaxies, leading to lower emission fluxes than we predict, and that the inner gaseous halos are slightly more massive or denser around star forming galaxies, leading to higher emission fluxes than we predict.  Perhaps most disappointing is that the models fail to qualitatively reproduce the reversal observed between the higher fluxes at small $r_p$ for high SFR galaxies and higher fluxes at large $r_p$ in low SFR galaxies. We had expected this behavior to be the result of correlations between SFR and environment, but the model does not confirm this hypothesis.  

These discrepancies warrant further investigation with more realistic models and could reflect problems in our simplified halo gas characterizations or in how galaxies with certain SFR or morphology are assigned to halos. In either case, further study should be illuminating.  

\section{Conclusions}

Using over 8 million spectra that intersect the halos of
nearly half a million nearby galaxies, we find that the radial profile of H$\alpha$+N[{\small II}] emission line around normal galaxies extends well beyond 100 kpc in projected radius and has an inflection at a projected radius of $\sim$ 50 kpc. 

We also find trends in the halo emission line profiles with properties of the primary galaxy. At small projected radius, $r_p \sim $ 10 kpc, the line emission flux in the halos of high SFR or disk dominated galaxies is a few times higher than that in the low SFR or spheroid dominated galaxies. At  $r_p \sim 50$ kpc the emission flux profiles become somewhat more similar and 
at even larger $r_p$, $\sim 300$ kpc, the behavior reverses, with the high SFR and disk dominated galaxies having lower levels of emission. These results are for stellar mass-matched samples. 
When we control for both stellar mass and morphology, we still find differences in the halo emission that are related to the star formation rate of the central galaxy. These results highlight the connectivity between the central galaxy and the surrounding CGM, extending beyond its immediate vicinity.
The emission profile, and its dependence on galaxy properties, promises to be a key constraint on models of the CGM/galaxy connection. 

To further understand the emission profile inflection and relationships between the primary galaxy and the 
emission profile, we constructed a simple, but cosmologically motivated, halo model, where each halo is populated with gas using a hydro-equilibrium isothermal NFW profile. The gas is 
split into a hot component at the virial temperature and a cooler component, T = 12,000 K. This model
matches the emission profile from the full sample extremely well whether we fit the fraction of cold gas in primary and secondary halos (0.26 and 0.34) or we adopt values published previously derived from hydrodynamical simulations (0.19 and 0.38). We demonstrated statistically that the 
contribution from associated
halos are necessary to produce the excellent match. At $r_p > 50$ kpc, the emission flux is primary coming from these  neighboring halos. This result explains both the inflection in the radial emission line profile and the decreasing correlation we find between the emitted flux and the properties of the primary galaxy for increasing $r_p$. 

Our model is manifestly oversimplified as we ignore temperature variations, assume a smoothly varying cold gas density profile, and procedurally assign galaxies to dark matter halos. As such, perhaps it is not too surprising that despite the successes described above
the model fails to reproduce more detailed observational findings that we discuss. This failure is likely indicating that the cold gas properties (fraction, temperature, density) depend on aspect of galaxy formation and evolution that have not been included in our simple model. 
Either way, additional, more realistic modeling of the halo emission line fluxes has the potential to help us understand important aspects of galaxy structure and evolution. We provide our measurements as a constraint for future theoretical models.

\section{Acknowledgments}
DZ and HZ acknowledge financial support from NASA ADAP NNX12AE27G and NSF grant AST-1311326. The authors gratefully acknowledge Yinzhe Ma and Houjun Mo for helpful discussions, and the SDSS III team for providing a valuable resource to the community.

Funding for SDSS-III has been provided by the Alfred P. Sloan Foundation, the Participating Institutions, the National Science Foundation, and the U.S. Department of Energy Office of Science. The SDSS-III web site is http://www.sdss3.org/.

SDSS-III is managed by the Astrophysical Research Consortium for the Participating Institutions of the SDSS-III Collaboration including the University of Arizona, the Brazilian Participation Group, Brookhaven National Laboratory, Carnegie Mellon University, University of Florida, the French Participation Group, the German Participation Group, Harvard University, the Instituto de Astrofisica de Canarias, the Michigan State/Notre Dame/JINA Participation Group, Johns Hopkins University, Lawrence Berkeley National Laboratory, Max Planck Institute for Astrophysics, Max Planck Institute for Extraterrestrial Physics, New Mexico State University, New York University, Ohio State University, Pennsylvania State University, University of Portsmouth, Princeton University, the Spanish Participation Group, University of Tokyo, University of Utah, Vanderbilt University, University of Virginia, University of Washington, and Yale University.

\bibliography{bibliography}

\end{document}